\documentclass[12pt]{article}
\usepackage{epsf}
\def\be{\begin{equation}}
\def\ee{\end{equation}}
\def\bea{\begin{eqnarray}}
\def\eea{\end{eqnarray}}

\begin{document}
\begin{titlepage}
\begin{center}
{\Large \bf William I. Fine Theoretical Physics Institute \\
University of Minnesota \\}
\end{center}
\vspace{0.2in}
\begin{flushright}
FTPI-MINN-18/14 \\
UMN-TH-3725/18 \\
August 2018 \\
\end{flushright}
\vspace{0.3in}
\begin{center}
{\Large \bf Quark spin structures of bottom-charmed threshold molecular states.
\\}
\vspace{0.2in}
{\bf  M.B. Voloshin  \\ }
William I. Fine Theoretical Physics Institute, University of
Minnesota,\\ Minneapolis, MN 55455, USA \\
School of Physics and Astronomy, University of Minnesota, Minneapolis, MN 55455, USA \\ and \\
Institute of Theoretical and Experimental Physics, Moscow, 117218, Russia
\\[0.2in]

\end{center}

\vspace{0.2in}

\begin{abstract}
The implementation of the heavy quark spin symmetry in possible molecular states at the mixed bottom-charm threshold is somewhat different from that at the open charm or bottom thresholds. In particular it depends on two parameters describing separately the symmetry for the bottom and for charmed quarks. The corresponding spin structures of the $S$-wave molecular states of the meson pairs $B^{(*)} D^{(*)}$ are discussed as well as their consequences for properties of possible near-threshold resonances, analogs of the known charmonium-like and bottomonium-like $X$ and $Z$ states. 
  \end{abstract}
\end{titlepage}

Charmonium-like and bottomonium-like states with masses very near the open heavy flavor thresholds reveal an interesting and intricate strong-interaction dynamics. Some of these states are very likely related to the dynamics of heavy meson-antimeson pairs, such as e.g. the $Z_b(10610)$ and $Z_b(10650)$ resonances~\cite{bellez} at the respective thresholds of $B \bar B^*$ and $B^* \bar B^*$ pairs, or their charmonium-like analogs $Z_c(3900)$~\cite{besz39} and $Z_(4020)$~\cite{besz40}, as well as the `peak that started it all' $X(3872)$~\cite{pdg} at the very threshold of $D^{*0} \bar D^0$ pair of charmed mesons. (Recent reviews of these and other exotic resonances in the charmonium-like and bottomonium-like sectors can be found in Refs.~\cite{Guo17} and \cite{Ali17}.) In particular, the decay properties of the $Z_b$ resonances, namely their comparable decay rate to final states $\pi  \Upsilon(nS) $ with $n=1$, 2 and 3 and to $\pi  h_b(kP) $ with $k=1$ and 2, indicates that the structure of the heavy and light spin states in these resonances follows that in free $S$-wave meson-antimeson pairs~\cite{bgmmv}. The spin signature is revealing of the internal structure due to the relative weakness of the QCD interaction depending on the spin of a heavy quark, which interaction is inversely proportional to the heavy mass, thus implying an (approximate) conservation of this spin --- the so-called heavy quark spin symmetry (HQSS). Indeed, in widely separated mesons the spin of the heavy quark (antiquark) is correlated with the spin of the light antiquark (quark). On the oter hand in the heavy quarkonium  the spins of the heavy quark and antiquark correlate with each other and combine into the total spin, $S_{b \bar b} =1$ in  $\Upsilon(nS)$ and $S_{b \bar b} =0$ in $h_b(kP)$. The appearance in the decays of the states with different $S_{b \bar b}$ is due to the comparable (essentially equal) presence of both final-state spin states in the expansion of the initial spin wave function in the eigenstates of the total spin of the heavy (and the light) quark-antiquark pair~\cite{bgmmv}. Furthermore, by considering other HQSS partners of the observed states, one can expect existence of other four-quark near-threshold resonances which thus far escaped observation due to their different quantum numbers~\cite{mv11}.

It can be noted that for the meson pairs with a hidden heavy flavor, i.e. $D^{(*)} \bar D^{(*)}$ or $B^{(*)} \bar B^{(*)}$, there are additional implications  for the heavy-light spin structure following from the charge symmetry for the neutral pairs, or from an approximate isospin symmetry for the charged ones. In particular, the $C$-even pair, e.g. $D^0 \bar D^{*0} + \bar D^0 D^{*0}$, such as in the resonance $X(3872)$, necessarily contains the heavy $c \bar c$ pair in the spin state $S_{c \bar c}=1$ and the light degrees of freedom also have the total spin $S_L$ equal to one~\cite{mv04}, $S_L=1$, and the two total spins combine into the overall angular momentum $J=1$:  $D^0 \bar D^{*0} + \bar D^0 D^{*0} \sim (1_{c \bar c} \otimes 1_L) |_{J=1}$. On the other hand, the $C$-odd combination is an equal mixture of spin states: e.g. $B \bar B^* - B^* \bar B \sim (0_{b \bar b} \otimes 1_L) + (1_{b \bar b} \otimes 0_L)$ as is the case for the $Z_b(10610)$ resonance. For the charged meson-antimeson combinations the separation of the symmetric and antysimmetric states is ensured by the $G$ parity (i.e. the $C$ and the isospin symmetry). The required for such separation degeneracy of the conjugate states, e.g. $D^{*+} \bar D^0$ and $D^+ \bar D^{*0}$, is not quite exact due to a slightly different mass splitting in the isotopic doublets of the pseudoscalar $D$ and vector $D^*$ mesons and in principle may give rise to small but observable effects in the charged $Z_c(3900)$ peak~\cite{mv16}.

The spin structure of the near-threshold heavy resonances is directly related to another intriguing question of the dynamics of these states. Namely, the question is what forces between the heavy mesons are responsible for the formation of the threshold structures? The long-range interaction mediated by one-pion exchange~\cite{mp,nv}, or by an $\eta$ exchange~\cite{kr}, appears~\cite{mv16l,bonn18} not to be the major factor in the dynamics of heavy meson pairs. However the issue of the significance of these forces remains open.

The purpose of the present paper is to point out that some of the questions of molecular dynamics of heavy mesons can be studied in a somewhat different setting if in addition to charmonium-like and bottomonium-like states one could also study mixed heavy flavor systems containing $c$ quark and $\bar b$ antiquark. It is needless to mention that unlike for the charmonium-like and bottomonium-like states, accessible in the $e^+e^-$ annihilation, there is no comparable clean source of the mixed heavy flavor systems, with high energy hadronic colliders being the only available platform for studying bottom-charm objects. For this reason the mixed-flavor quarkonium has been studied to a much lesser extent with only the $B_c(1S)$ and $B_c(2S)$ mesons observed thus far. One can speculate however that future development of experimental technique, e.g. along the lines of the LHCb experiment, may make feasible studies of other states of the $\bar b c$ quarkonium as well as of two-meson states at the thresholds of $B^{(*)} D^{(*)}$. It thus appears worthwhile to (temporarily) put aside the legitimate doubts about practical feasibility of experimental studies and to discuss possible advantages of studying the mixed heavy flavor systems.   

In this paper we follow the nomenclature adopted by PDG~\cite{pdg}, and call $D$ mesons those containing the charmed quark, $D \sim c \bar q$, and $B$ mesons  those containing the bottom antiquark, $B \sim \bar b q$. The meson pairs  $B^{(*)} D^{(*)}$ then have the same bottom-charm quantum numbers as the $B_c$ mesons, $B_c \sim {\bar b} c$, and are those discussed in this paper. Unlike the systems with two heavy quarks (e.g. $bc$~\cite{kr2,eq}) those with the considered here quantum numbers always have a $\bar b c$ quarkonium state [plus possibly light meson(s)] as the lowest mass state, so that any peaks near  two-meson thresolds are necessarily resonances.  It will be argued here that studies of possible near-threshold molecular resonances in two-meson systems may provide additional data on the forces between the mesons. In particular, it can be readily noted that in neither of such pairs, except for $B^* D^*$, an exchange of a pseudoscalar meson (e.g. $\pi$ or $\eta$) is possible, so that existence or non-existence of threshold resonances can be studied without a `contamination' from the effects of such exchanges. Furthermore, the thresholds in the channels $B^* D$ and $B D^*$ are strongly split in mass (by about 95\,MeV) so that the effects of the `spin exchange' between the mesons can be well neglected.

Due to HQSS the spin structure of the threshold states reveals itself in their transitions to the states of quarkonium with emission of light mesons, as is conspicuously demonstrated~\cite{bgmmv} by the transitions from the $Z_b$ resonances to the states of bottomonium. In order to discuss this behavior for the mixed $\bar b c$ systems one should take into account a well known in the literature significant difference in the spin counting for the $\bar b c$ quarkonium from that in bottomonium or charmonium. Namely, the states of the latter systems naturally fall into either para- or ortho- category with the total spin of the heavy quark-antiquark pair being respectively one or zero. The complete separation between these states is ensured by that they are also eigenstates of the $C$ parity. This is not the case for the mixed $\bar b c$ systems, so that the states of this system do not have to have definite total spin of the $\bar b c$ pair, unless required by the overall angular momentum conservation. In particular the $S$-wave states $^1S_0$ and $^3S_0$ have to be pure spin states, since there is no orbital momentum at all in these. Other pure spin (ortho-) states are the $^3P_2$ and $^3P_0$ (with $J^P$ equal to $2^+$ and $0^+$ respectively), since $S_{\bar b c}=1$ is required to combine with the orbital momentum $\ell =1$ in the required overall angular momentum. The situation however is different for the axial states with $J^P=1^+$, which in fact can be mixed ortho- and para- states, $^3P_1$ and $^1P_1$~\cite{eq94}. 

In the limit where the bottom quark is infinitely heavy, so that its spin completely decouples, it is the total angular momentum $\vec j = \vec S_c + \vec \ell$ of the orbiting charmed quark that would be conserved. For a $P$-wave level this can take values $j=1/2$ and $j=3/2$ making the spin-parity of the states of the $c$ quark as $(1/2)^-$ and $(3/2)^-$. Clearly, when combined with the $(1/2)^-$ spinor of the $\bar b$ antiquark,  the resulting $J^P=1^+$ states can be written as mixed $^3P_0$ and $^1P_1$ states~\cite{eq94}: 
\bea
P_1 \equiv \left . \left [ \bar b \otimes \left ( {1 \over 2} \right )^- \right ] \right |_{J=1} & = &  {\sqrt{2 \over 3}} \, \left |  ^3P_1 \right \rangle+ {\sqrt{1 \over 3}} \, \left | ^1P_1 \right \rangle  \nonumber \\
P_1^{'} \equiv \left . \left [ \bar b \otimes  \left ( {3 \over 2} \right )^- \right ] \right |_{J=1} & = & - {\sqrt{1 \over 3}} \, \left |  ^3P_1 \right \rangle+ {\sqrt{2 \over 3}} \, \left | ^1P_1 \right \rangle,
\label{pmix}
\eea
corresponding to the mixing angle $\theta=\arctan({1/\sqrt{2}}) \approx 35^\circ$. In this picture the $^3P_0$ state of the quarkonium is $[\bar b \otimes (1/2)^-]|_{J=0}$ and forms a doublet of the $\bar b$ spin symmetry with the $P_1$, while the remaining $^3P_2$ state is $[\bar b \otimes (3/2)^-]|_{J=2}$ and makes a doublet with the $P_1^{'}$.

For realistic heavy quarks one should expect a deviation from the idealized picture of Eq.(\ref{pmix}) described by the parameter $m_c/m_b \approx 0.3$. Indeed, in the available calculations within potential models~\cite{gklt,gj,Fulcher,efg,Godfrey}, the mixing angle comes out to be in the range $20^\circ - 30^\circ$ and is in the ballpark of the expected deviation from the ideal value. 
In either case it appears to make sense to consider as a starting point the approximation of an infinitely heavy bottom quark as well as the limit of HQSS for the charmed quark. The parameter for the latter approximation is $\Lambda_{QCD}/m_c$ whose value is numerically comparable to the parameter $m_c/m_b$ for the former one. It is the double limit of both parameters being negligibly small that is assumed throughout the rest of this paper.

As is mentioned, for a discussion of transitions from the two-meson states to those of the ${\bar b} c$ quarkonium with a definite total spin $S_{\bar b c}$, it is necessary to consider 
the expansion of the $S$ wave states of meson pairs in terms of eigenstates of the total spin $S_{\bar b c}$ of the heavy $\bar b c$ pair and the total spin $S_L$ of the light ${\bar q} q$ pair. This expansion can be  essentially read off the known formulas~\cite{mv11} for the hidden-bottom meson-antimeson pairs, with the only difference being that the $B^* D$ and $B D^*$ are analogs of the sum and the difference of the definite charge parity states  $B^* \bar B \pm B \bar B^*$. The expansion for the molecular states in the ascending order of the threshold energy can thus be written as
\be
 B D, ~~7145\,{\rm MeV},~~0^+: ~~~{1 \over 2}\, \left ( 0^-_{\bar b c} \otimes 0^-_L \right ) - {\sqrt{3} \over 2} \, \left .\left ( 1^-_{\bar b c} \otimes 1^-_L \right ) \right |_{J=0}~;
\label{bd}
\ee
\be
B^* D, ~~7190\,{\rm MeV},~~1^+: ~~~{1 \over 2} \, \left ( 0^-_{\bar b c} \otimes 1^-_L \right ) + {1 \over 2} \, \left ( 1^-_{\bar b c} \otimes 0^-_L \right ) + {1 \over \sqrt{2}} \, \left .\left ( 1^-_{\bar b c} \otimes 1^-_L \right ) \right |_{J=1}~;
\label{bsd}
\ee
\be
B D^*,~~7285\,{\rm MeV},~~1^+: ~~~{1 \over 2} \, \left ( 0^-_{\bar b c} \otimes 1^-_L \right ) + {1 \over 2} \, \left ( 1^-_{\bar b c} \otimes 0^-_L \right ) - {1 \over \sqrt{2}} \, \left .\left ( 1^-_{\bar b c} \otimes 1^-_L \right ) \right |_{J=1}~;
\label{bds}
\ee
\be
\left . B^* D^*, ~~7330\,{\rm MeV}, \right |_{J=0}, ~~0^+: ~~~ {\sqrt{3} \over 2}\, \left ( 0^-_{\bar b c} \otimes 0^-_L \right ) + {1 \over 2} \, \left .\left ( 1^-_{\bar b c} \otimes 1^-_L \right ) \right |_{J=0}~;
\label{bsds0}
\ee
\be
\left . B^* D^* \right |_{J=1}, ~~1^+: ~~~{1 \over 2} \, \left ( 0^-_{\bar b c} \otimes 1^-_L \right ) - {1 \over 2} \, \left ( 1^-_{\bar b c} \otimes 0^-_L \right )~;
\label{bsds1}
\ee
\be
\left . B^* D^* \right |_{J=2}, ~~2^+: ~~~\left .\left ( 1^-_{\bar b c} \otimes 1^-_L \right ) \right |_{J=2}~,
\label{bsds2}
\ee
where the approximate mass of the corresponding threshold and the $J^P$ quantum numbers are also indicated for each channel.

The expansion in Eqs.(\ref{bd} - \ref{bsds2}) suggests certain relations between the amplitudes of transitions from the two-meson pairs to the $\bar b c$ quarkonium. In particular, the formula (\ref{bsd})  implies that an $S$-wave state of the pair $B^*D$ contains both the $0^-_{\bar b c} \otimes 1^-_L $ and $1^-_{\bar b c} \otimes 1^-_L $ components. Due to HQSS in a transition to the $S$ states of the $\bar b c$ quarkonium, the $0_{\bar b c}$ state goes into the pseudoscalar $^1S_0$, while the $1^-_{\bar b c}$ goes (with the same amplitude) to the vector one $^3S_1$. The state of the light degrees of freedom $1^-_L$ is the same in both transitions. Comparing the coefficients of the two components in Eq.(\ref{bsd}), one can thus expect the following relation between the decay rates
\be
{\Gamma[(B^* D) \to B_c(^3S_1) + {\rm light~ meson}] \over \Gamma[(B^* D) \to B_c(^1S_0) + {\rm light~ meson}]} \approx 2~,
\label{3s1s}
\ee
where the ``light meson'' can be either $\rho$ or $\omega$, depending on the isospin of the $B^* D$ meson pair. Clearly, the same relation follows from Eq.(\ref{bds}) for the heavier $B D^*$ pairs. It can be noticed that emission of a single light pseudoscalar meson, $\pi$ or $\eta$, is possible only in the transition of a $B^*D$ or $B D^*$ pair to only the $^3S_1$ quarkonium, since a transition to the singlet $0^-$ state is forbidden by parity. Certainly, the formulas (\ref{bsd}) and (\ref{bds}) are in full agreement with this requirement.

In terms of the approach where the masses of the charmed and bottom quarks are treated as `diffently heavy' the strongest spin dependent interaction among the constituents $\bar b c \bar q q$ of the meson pair is still that of the light quark-antiquark pair, so that $S_L$ still appears to be a good quantum number. This interaction is followed in strength by that of the charmed quark with the spin interaction of the $\bar b$ being the weakest. Accordingly, one can consider combining the spin of the $c$ quark with that of the light pair into the angular momentum eigenstates:
\bea
&& \chi, ~~ \left ( {1 \over 2} \right )^-: ~~~c \otimes 0^-_L~; \nonumber \\
&& \phi, ~~ \left ( {1 \over 2} \right )^-: ~~~\left. \left ( c \otimes 1^-_L \right ) \right |_{J=1/2}~; \nonumber \\
&& \psi, ~~ \left ( {3 \over 2} \right )^-: ~~~\left. \left ( c \otimes 1^-_L \right ) \right |_{J=3/2}~,
\label{cpp}
\eea
where $c$ stands for the spinor of the charmed quark, so that its quantum numbers are $J^P=(1/2)^+$.
The expansion of the states of meson pairs in terms of combinations of these eigenstates with the spin states of the $\bar b$ quark ($J^P=(1/2)^-$) can be readily found and reads as
\be
 B D,~~0^+: ~~~ {1 \over 2} \, \left. \left ( {\bar b} \otimes \chi \right ) \right |_{J=0} - {\sqrt{3} \over 2} \, \left. \left ( {\bar b} \otimes \phi \right ) \right |_{J=0}~;
\label{fbd}
\ee
\be
B^* D,~~1^+: ~~~{1 \over 2} \, \left. \left ( {\bar b} \otimes \chi \right ) \right |_{J=1} + {\sqrt{3} \over 2} \, \left. \left ( {\bar b} \otimes \phi \right ) \right |_{J=1}~;
\label{fbsd}
\ee
\be
B D^*,~~1^+: ~~~\sqrt{2 \over 3} \, \left. \left ( {\bar b} \otimes \psi \right ) \right |_{J=1}-{1 \over 2} \, \left. \left ( {\bar b} \otimes \chi \right ) \right |_{J=1} + {1 \over 2 \sqrt{3}} \, \left. \left ( {\bar b} \otimes \phi \right ) \right |_{J=1}~;
\label{fbds}
\ee
\be
\left . B^* D^* \right |_{J=0}, ~~0^+: ~~~  {\sqrt{3} \over 2} \, \left. \left ( {\bar b} \otimes \chi \right ) \right |_{J=0} + {1 \over 2} \, \left. \left ( {\bar b} \otimes \phi \right ) \right |_{J=0}~;
\label{fbsds0}
\ee 
\be
\left . B^* D^* \right |_{J=1}, ~~1^+: ~~~ - {1 \over \sqrt{3}} \, \left. \left ( {\bar b} \otimes \psi \right ) \right |_{J=1}-{1 \over \sqrt{2}} \, \left. \left ( {\bar b} \otimes \chi \right ) \right |_{J=1} + {1 \over  \sqrt{6}} \, \left. \left ( {\bar b} \otimes \phi \right ) \right |_{J=1}~;
\label{fbsds1}
\ee
\be
\left . B^* D^* \right |_{J=2}, ~~2^+: ~~~\left. \left ( {\bar b} \otimes \psi \right ) \right |_{J=2} ~.
\label{fbsds2}
\ee
It can be pointed out that the asymmetry between the expressions (\ref{fbsd}) for the $B^*D$ channel and (\ref{fbds}) for $B D^*$ is a direct result of different treatment of the spins of the ${\bar b}$ and $c$ quarks due to a `two-tier' hierarchical implementation of HQSS.

An application of the expansion in Eqs.(\ref{fbd} - \ref{fbsds2}) arises in considering the processes involving the molecular two-meson states and the $P$ wave $\bar b c$ quarkonium. One such process is the mixing of molecular and quarkonium states that is allowed for isotopically singlet two meson states. In particular, the mixing of the two molecular $J^P=1^+$ states, i.e. 
\be
X_{\bar b c} = {B^{*+} D^0 + B^{*0} D^+ \over \sqrt{2}}~~~ {\rm and}~~~ X_{\bar b c}^{'} = {B^+ D^{*0} + B^0 D^{*+} \over \sqrt{2}}~,
\label{xxp}
\ee
should involve different combinations of the quarkonium states $P_1$ and $P_1^{'}$. Namely, according to Eq.(\ref{fbds}) the $B^*D$ pair is allowed to mix only with the $P_1$ quarkonium, while the $B D^*$ pairs generally mix with both $P_1$ and $P_1^{'}$ states. It can be mentioned in connection with the mixing that the first radially excited $P_1$ state, $2P_1$, of ${\bar b c}$ is expected~\cite{gklt,eq94,zvr,efg,Godfrey} to have mass between approximately 7124 and 7150\,MeV. This expected mass is quite close to the threshold for the $B D^*$ pairs (at approximately 7190\,MeV) and the mixing with a possible molecular threshold state $X_{\bar b c}$ can be strong. The strength of the mixing with the heavy quarkonium is related to the topic of production of the threshold state in hard processes, e.g. in high energy proton collisions at the LHC. Indeed, it has been pointed out~\cite{bk} in the course of the studies of the charmonium-like state $X(3872)$ that a shallow bound molecular state would be only very weakly produced in such hard processes as $B$ decays or in proton-antiproton collisions, where the $X(3872)$ has been observed. The production can be explained if there is an admixture of a spatially compact $^3P_1$ charmonium component in the wave function of $X(3872)$, which is the component produced in hard processes. (A discussion of this point can be found e.g. in the review \cite{mvch}.) Since any possible future studies of the discussed bottom-charm states would necessarily run into the problem of finding sources for their production, it appears most promising that the state $X_{\bar b c}$ in Eq.(\ref{xxp}) can be produced at some measurable rate in high energy collisions due to the mixing with compact $\bar b c$ quarkonium. The molecular states that cannot mix with quarkonium, such as all those with nonzero isospin, essentially are not produced in hard collisions, and in principle can be observed in `soft' transitions from higher isosinglet states (analogously to e.g. the transitions $\Upsilon(5S) \to Z_b \pi$, which serve as a source of the $Z_b$ resonances). However such production mechanism subjects the prospects of observing isotopically non-singlet bottom-charm molecular states to an even greater uncertainty of unknown properties of unknown higher bottom-charm resonances.

It should be mentioned that in terms of the proximity of masses of the mixing states a similarly advantageous situation is with the $BD$ threshold state and the $2^3P_0$ quarkonium, which states are allowed to mix according to Eq.(\ref{bd}). A search for a peak in this channel with $S$-wave decays to either $B_c(^1S_0) \eta$ or $B_c(^3S_1) \omega$ may reveal or exclude existence of a threshold $BD$ resonance, which would be of a great interest given that $J^P=0^{++}$ resonances made of $D \bar D$ or $B \bar B$ have not yet been observed.  

A study of the isotopic properties of the state $X_{\bar b c}$ may shed light on another question of heavy meson dynamics related to $X(3872)$. Namely, the peripheral part of the wave function of $X(3872)$ contains mostly the neutral charmed mesons: $D^0 \bar D^{*0}+ D^{*0} \bar D^0$, which is a mixture of isotopic singlet and a triplet. This is attributed to the fact that the threshold for the other isotopically related component $D^+ D^{*-}+ D^{*+} D^-$ is heavier by about 8\,MeV, which is a substantial energy gap in the scale of of the splitting between the $X(3872)$ mass and the $D^0 \bar D^{*0}$ threshold. Moreover, the interaction between the heavier component and the lighter state can `push' the latter down in mass and may contribute to the very existence of the threshold peak. The significance of the isotopic mass splitting of the mesons is apparently different in the $B^* D$ system, since this splitting is small for the $B^*$ mesons and it is the $4.8$\,MeV mass gap between $D^+$ and $D^0$ that mostly contributes to a a possible isospin violation in $X_{\bar b c}$. It is not clear at present whether such smaller, than in $X(3872)$, amount of isospin violation would prevent an appearance of the threshold resonance $X_{\bar b c}$. If this state exists as a peak, a measurement of the ratio of its decay rates $\Gamma(X_{\bar b c} \to B_c \rho)/\Gamma(X_{\bar b c} \to B_c \omega)$ can serve as a gauge of the amount of the isospin violation. It can be noted that in comparison with $X(3872)$ this ratio of the transition rates is much less affected by the kinematically available phase space,  since the mass difference between the heavy states is about 915\,MeV.

A related suite of processes involving the discussed bottom-charm meson pairs and the quarkonium $P$-wave states are the transitions with emission of an $S$-wave pair of pions, e.g. $X_{\bar bc} \to 1P \pi \pi$. The $1P$ states are expected to be below 6800\,MeV, so that such transitions are fully kinematically allowed. In particular Eq.(\ref{fbsd}) tells us that in the limit of HQSS the transitions of this type from $X_{\bar b c}$ to only the $1P_1$ state are allowed, $X_{\bar b c} \to 1P_1 \pi \pi$ while for the heavier $X_{\bar b c}^{'}$ both $1P_1$ and $1P_1^{'}$ are allowed to be present in the final state. An observation of such transitions would also be interesting from the point of possible comparison with $X(3872)$, for which similar pion transitions to $P$-wave charmonium were discussed~\cite{dv}, but not yet observed. 

The discussed considerations illustrate that a study of differences and similarities between the threshold bottom-charm states and those in the charmonium-like and bottomonium-like sectors may greatly contribute to understanding of the dynamics of heavy mesons and shed light on the general issue of the `XYZ' states. The presence of two parameters, $\Lambda_{QCD}/m_c$ and $m_c/m_b$ describing the implementation and the violation of HQSS for separately the charmed and the $b$ quarks makes the treatment of the spin structures of the molecular states different from the cases where heavy quarks of only one flavor are present. In particular this results in that two types of expansion of the spin wave function of the heavy meson pairs are necessary [Eqs.(\ref{bd} - \ref{bsds2}) and Eqs.(\ref{bd} - \ref{bsds2})], each expansion being of use for considering separate class of processes --- such as those involving ${\bar b c}$ quarkonium with definite total spin of the heavy quark pair and those where the quarkonium states have the spin of the $c$ quark and the orbital momentum combined in (approximately) definite eigenstates. An observation of the mixed heavy flavor molecular states is undoubtedly very challenging and it is not clear at present whether such observation may become feasible at all. One can recall however that it took quite some time between the first observation of charmonium in 1974 and the discovery of the first (very likely) molecular state $X(3872)$ in 2003. The $B_c$ meson, the lowest state of the ${\bar b c}$ quarkonium, was first observed in 1998. Thus it may be that it is just a matter of time and development of the experiments before it would become possible to study experimentally bottom-charm molecular states. The details of the hadronic dynamics that such sytems allow to access may be well worth the effort of trying to reach them in experiments.

This work is supported in part by U.S. Department of Energy Grant No.\ DE-SC0011842.

\end{document}